\documentclass[letterpaper,10pt]{article}   
\usepackage{osajnl2} 
\usepackage{amsmath}

\begin{document}

\title{A Sampling Theorem for Computational Diffraction}
\author{Daniel J. Merthe$^{1,*}$}

\address{$^1$University of Southern California, \\ Los Angeles, CA 90089, USA}
\address{$^*$Corresponding author: merthe@usc.edu}

\begin{abstract}
A major challenge of many diffraction calculations, using some form of the Rayleigh-Sommerfeld formulas, is the integration of a highly oscillatory integrand. Here we derive a potentially useful alternative form of solution to the Helmholtz equation, which implies a sampling theorem for the evaluation of a diffracted scalar field. This alternative solution bears close resemblance to the Rayleigh-Sommerfeld diffraction formulas, but instead incorporates the boundary conditions digitally. Hence, the integration is replaced by a simple summation. This formulation may be more efficient for accurate computer-based calculation of the diffracted scalar field.
\end{abstract}

\date{}                                           

\maketitle
\section{Introduction}
The diffraction of monochromatic electromagnetic radiation by obstacles much larger than the wavelength is often treated satisfactorily by the Scalar Diffraction Theory. That is, one solves the Helmholtz equation with certain boundary conditions. We let $U$ be any component of either the electric or magnetic fields, defined in the half-space $z \geq 0$, free of charges or currents. If the scalar field $U$ is simple harmonic in time with frequency $\omega$, then it satisfies the Helmholtz equation,
\begin{equation}
\nabla^2 U + k^2 U = 0
\label{Helmholtz}
\end{equation}
where $\nabla^2$ is the Laplacian differential operator, $k = \omega / c = 2 \pi / \lambda$ is the wavenumber, $c$ is the speed of light and $\lambda$ is the wavelength. The boundary conditions are usually given as follows \cite{Goodman:1988fk}. For some known function $U_0(x, y)$, the scalar field $U$ on the boundary plane $z = 0$ is $U(x, y, 0) = U_0 (x, y)$. If $r$ is the distance from the origin to any point in the half-space, we assume the Sommerfeld radiation condition, $\lim_{r \rightarrow \infty} U = O(1/r)$, such that the total energy is bounded. Additionally, we assume that all radiation is propagating in the positive $z$-direction.

One solution to this boundary value problem is the Rayleigh-Sommerfeld diffraction formula of the first kind, which can be written as \cite{Goodman:1988fk},
\begin{equation}
U(x,y,z) = \frac{-1}{2 \pi} \iint U_0 (x',y') \frac{\partial}{\partial z} \frac{\exp (i k R)}{R} dx' dy'
\label{Rayleigh}
\end{equation}
where $R$ is the distance between the points $P$ and $Q$ with coordinates $(x, y, z)$ and $(x', y', 0)$, respectively, and the integration is taken over the entire plane $z = 0$. This integral can be evaluated exactly only for a small class of boundary value functions $U_0(x, y)$. Under certain additional constraints, it can be approximated by the Fresnel or Fraunhofer diffraction integrals \cite{Born:1980uq,Goodman:1988kx,Hecht:2002fk}, which essentially reduces the problem to Fourier analysis. However, the diffracted field is often found from Eq. (\ref{Rayleigh}), given some general incident field $U_0(x,y)$, by evaluating the integral numerically using a computer.

The principal challenge of evaluating the right side of Eq. (\ref{Rayleigh}) numerically is creating a sampling mesh fine enough to accurately represent the highly oscillatory integrand. The exponential term oscillates spatially with a mimimum wavelength of $\lambda$. Thus, it seems necessary that the spacing between mesh points be much smaller than $\lambda$. Here we show that a sampling mesh with spacing $\lambda/2$ between points is sufficient to obtain high accuracy for most practical applications. Moreover, if the field is composed only of homogeneous plane waves, then this produces the exact result.

We propose a modified version of the Rayleigh-Sommerfeld solution, which incorporates the boundary conditions discretely. To arrive at this formulation, we employ the angular spectrum method \cite{Mandel:1995fk} of field analysis. Both the two-dimensional and three-dimensional versions of the result are discussed.

\section{The Angular Spectrum: Homogeneous and Inhomogeneous Wave Components}
The Rayleigh-Sommerfeld solution to the Helmholtz equation is obtained by using the Kirchhoff integral formula in conjunction with the Green's function solution \cite{Goodman:1988fk}. Yet, there is an alternative method for solving Eq. (1), referred to as the angular spectrum method. It can be shown\cite{Mandel:1995fk,Goodman:1988fk} that the general solution to Eq. (2) in the half-space $z \geq 0$ with the given boundary conditions can be written as
\begin{equation}
U(x,y,z) = \iint_{-\infty}^{\infty} A(u,v) \exp (ik [ ux + vy + wz ]) \ du \ dv
\label{AngSpec}
\end{equation}
where $w = +\sqrt{1-u^2-v^2}$, and
\begin{equation}
A(u,v) = \lambda^{-2} \iint_{-\infty}^{\infty} U_0(x,y) \exp (-ik [ux + vy]) \ dx \ dy
\end{equation}
is the so-called angular spectrum of the field $U$, as a function of the direction cosines $u$ and $v$. Evidently, the angular spectrum function is just the Fourier transform  of the boundary value function $U_0 (x,y)$. For each plane wave component,
\begin{equation*}
U_{u,v} = A(u,v) \exp ( ik [ux+vy+wz])
\end{equation*}
in the integrand of Eq. (\ref{AngSpec}), there are two physically distinct possibilities: $u^2+v^2 \leq 1$ and $u^2 + v^2 > 1$. In the former, $w$ is real and the plane wave is called homogeneous. In the latter case, $w$ is imaginary and the plane wave is called inhomogeneous,
\begin{equation}
\exp ( ik [ux+vy+wz]) = \exp ( - k |w| z) \exp ( ik [ux+vy]).
\end{equation}
This latter kind of wave is also known as an evanescent wave, because it propagates in the plane $z = 0$ and vanishes rapidly in the positive $z$ direction. Based on this consideration, we can express the field $U$ as the sum of its homogeneous and inhomogeneous plane wave components,
\begin{equation}
U(x,y,z) = U_H(x,y,z) + U_I(x,y,z)
\end{equation}
where the homogeneous component is
\begin{equation}
U_H(x,y,z) = \iint_{u^2+v^2 \leq 1} A(u,v) \exp (ik [ ux + vy + wz ]) \ du \ dv,
\label{homo}
\end{equation}
and the inhomogeneous component is
\begin{equation}
U_I(x,y,z) = \iint_{u^2+v^2 > 1} A(u,v) \exp ( -k |w| z) \exp ( ik [ux+vy]) \ du \ dv.
\label{inhomo}
\end{equation}
Let us place an upper bound on $| U_I |$ for any (real) value of $kz$. Applying Schwarz' inequality to the right side of Eq. (\ref{inhomo}) gives
\begin{equation}
\begin{split}
| \iint_{u^2+v^2 > 1} A(u,v) \exp ( -k |w| z) & \exp ( ik [ux+vy]) \ du \ dv |^2 \\
&\leq \iint_{u^2+v^2 > 1} |A(u,v)|^2  \ du \ dv \ \cdot \  \iint_{u^2+v^2 > 1} \exp ( -2k |w| z) \ du \ dv
\end{split}
\end{equation}
The second integral on the right side is easily evaluated. Switching from Cartesian to polar coordinates in the space of the direction cosines, with radial coordinate $\rho = \sqrt{u^2+v^2}$, this integral takes the form,
\begin{equation}
\iint_{u^2+v^2 > 1} \exp ( -2k |w| z) \ du \ dv = 2 \pi \int_{1}^{\infty} \exp (-2kz \sqrt{\rho^2 - 1}) \ \rho \ d \rho
\end{equation}
Then, using the substitution $t^2 = \rho^2 - 1$ yields
\begin{equation}
2 \pi \int_{1}^{\infty} \exp (-2kz \sqrt{\rho^2 - 1}) \ \rho \ d \rho = 2 \pi \int_{0}^{\infty} \exp (-2kz t) \ t \ dt
\end{equation}
Finally, applying integration by parts to the right side has the result,
\begin{equation}
\iint_{u^2+v^2 > 1} \exp ( -2k |w| z) \ du \ dv = \frac{\pi}{2 k^2 z^2}
\end{equation}
Combining this with Eq. (10), we obtain
\begin{equation}
|U_I(x,y,z)| \leq \frac{1}{k z}  \sqrt{ \frac{\pi}{2}\iint_{u^2+v^2 > 1} |A(u,v)|^2 \  du \ dv }.
\end{equation}
In other words, for any given angular spectrum function $A(u,v)$, the optical field is given asymptically by 
\begin{equation}
U(x,y,z) = U_H (x,y,z) + O \left( 1/kz \right).
\label{neglUI}
\end{equation}
as $kz \rightarrow \infty$. In most applications, the value of $kz$ is very large, e.g. $\sim 10^6$ for visible light. It follows that the inhomogeneous plane wave component $U_I$ is usually nearly zero in almost all of the half-space $z >0 $. This contribution will be neglected in the following analysis.

In dealing only with the homogenous plane wave component, it is desirable to express $U_H$ in terms of its values at the boundary $z = 0$. The function $U_H(x,y,0)$ can be expressed in simple terms of the known function $U_0(x,y)$. It follows from elementary Fourier analysis and the convolution theorem that
\begin{equation}
U_H(x,y,0) = U_0 (x,y) \ast \frac{J_1 (k \sqrt{x^2+y^2})}{\lambda \sqrt{x^2+y^2}}
\label{homo0}
\end{equation}
where $\ast$ is the convolution operator and $J_1$ is the first order Bessel function of the first kind. It is easy to see that as the wavelength $\lambda$ approaches zero, the second term in the convolution becomes the two dimensional Dirac delta function $\delta^2(x,y)$, with the property $U_0 \ast \delta^2 = U_0$. In regions where $U_0(x,y)$ varies with length scales much larger than $\lambda$, this implies $U_H(x,y,0) \approx U_0(x,y)$. This approximation will fail at the boundaries of the obstructing obstacle where the field has a discontinuity. However, if the size of the obstacle is much larger than the wavelength, then the field near this boundary will contribute negligibly to the diffracted field at points far away from the obstacle.

\section{Diffraction in Two Dimensions}
Many configurations of diffraction are two-dimensional in form, e.g. diffraction by an infinitely long slit. Then as a first case, let us assume that the boundary field $U_0(x,y) = U_0(x)$ on the plane $z = 0$ has no dependence on the $y$-coordinate. Then, from Eq. (4)  the angular spectrum can be written as
\begin{equation}
A(u,v) = \delta(v) A'(u)
\end{equation}
where $\delta$ is the (one dimensional) Dirac delta function. Inserting this into Eq. (7) then yields
\begin{equation}
U_H(x,y,z) = U_H(x,z)  = \int_{-1}^{1} A(u) \exp (ik [ ux + wz ]) \ du
\label{homo2D}
\end{equation}
where here $w=+\sqrt{1-u^2}$ and the dash is dropped from the angular spectrum function $A(u)$. As expected, the homogeneous component in $z \geq 0$ also has no $y$-dependence. The same argument obviously applies to $U_I$. 

To evaluate the integral in Eq. (\ref{homo2D}), the function $A(u)$ need only be defined on the interval $u \in [-1,1]$. Within this domain, let us expand $A(u)$ as the Fourier series,
\begin{equation}
A(u) = \sum_{m=-\infty}^{\infty} c_m \exp (-i \pi m u)
\end{equation}
which is always possible for any sufficiently well-behaved function $A(u)$ \cite{Boas:2006uq}. The expansion coefficients $c_m$ are found by the inversion formula,
\begin{equation}
c_m = \frac{1}{2} \int_{-1}^{1} A(u) \exp (i \pi m u) \ du.
\label{FourierCoeff}
\end{equation}
Comparing Eq. (\ref{FourierCoeff}) to Eq. (\ref{homo2D}), it is immediately apparent that
\begin{equation}
c_m = \frac{1}{2} U_H \left( \frac{m \lambda}{2},0 \right).
\end{equation}
Combining this relation with Eqs. (16) and (17) therefore has the result
\begin{equation}
U_H (x,z)  = \sum_{m=-\infty}^{\infty} U_H \left( \frac{m \lambda}{2}, 0 \right) G \left( x-\frac{m \lambda}{2}, z \right)
\label{discretehomo}
\end{equation}
where
\begin{equation}
G(x,z) = \frac{1}{2} \int_{-1}^{1} \exp (ik [ux + wz]) \ du.
\label{G2D}
\end{equation}
We can compare Eq. (\ref{discretehomo}) to Eq. (2). For the homogeneous component $U_H$ of the optical field $U$, we replace the integral of Eq. (\ref{Rayleigh}) with the simple sum in Eq. (\ref{discretehomo}). That is, instead of requiring knowledge of the whole function $U_H$ on the boundary $z=0$, we only need to know its values at uniformly spaced points $x = m \lambda / 2$ for all integer $m$. Moreover, if the field is only non-zero for a finite region on $z=0$, then one only needs a finite set of known values of $U_H$ on $z=0$ to perfectly compute $U_H$ anywhere else in the half-space $z\geq 0$.

Let us now evaluate the integral on the right side of Eq. (\ref{G2D}) for a clearer expression of the kernel function $G(x,z)$. Firstly, consider the two important limiting cases: $z = 0$ and $z >> \lambda$. The first case is simple,
\begin{equation}
G(x,0) = \frac{1}{2} \int_{-1}^{1} \exp (ik ux) \ du = \sin (kx) / kx,
\end{equation}
which reduces the right side of Eq. (\ref{discretehomo}) to Whittaker's cardinal series expansion \cite{Whittaker:1915kx,Whittaker:1927vn,Shannon:1948uq,Goodman:1988fj} of the function $U_H (x,0)$. For the second case ($z >> \lambda$, or equivalently $kz>>1$), we make the trigonometric substitution, $u = \sin \eta$, such that $w = \cos \eta$ and using basic trigonometric identities Eq. (\ref{G2D}) becomes
\begin{equation}
G(x,z) = \frac{1}{2} \int_{-\pi/2}^{\pi/2} \exp (ik r \cos[\eta - \theta]) \ \cos \eta \ d \eta,
\end{equation}
switching to polar coordinates: $x = r \sin \theta$ and $y = r \cos \theta$. The restriction $z \geq 0$ implies that the angle $\theta$ is confined to $\theta \in [-\pi/2,\pi/2]$. Because we have $kz >> 1$, the exponential term oscillates rapidly, except near the stationary points of ${\cos [\eta - \theta]}$. These stationary points are $\eta = \theta + l \pi$ for any integer $l$. However, within the integration interval $[-\pi/2,\pi/2]$ there can only be one stationary point, namely $\eta = \theta$. Furthermore, near this point $\cos [\eta - \theta] \approx 1 - [\eta - \theta]^2/2$. Invoking the principle of stationary phase \cite{Erdelyi:1956zr}, it follows that
\begin{equation}
\frac{1}{2} \int_{-\pi/2}^{\pi/2} \exp (ik r \cos[\eta - \theta]) \ \cos \eta \ d \eta \approx \frac{\exp (ik r)}{2} \cos \theta   \int_{-\infty}^{\infty} \exp \left(- ik r \frac{[\eta - \theta]^2}{2} \right) \ d \eta.
\end{equation}
This well known integral is easily evaluated, resulting in the approximation
\begin{equation}
G(x,z>>\lambda) \approx \sqrt{\frac{\pi}{2}} \frac{ \exp (ik r)}{\sqrt{ikr}} \cos \theta.
\label{farfieldG}
\end{equation}
This function is just the asymptotic form (for large $kr$) of the two dimensional field established by an oscillating dipole at the origin, with the dipole moment aligned with the $x$-axis. Combining Eqs. (\ref{discretehomo}) and (\ref{farfieldG}), we obtain
\begin{equation}
U_H (x,z>>\lambda)  \approx  \sqrt{\frac{\pi}{2 i k}} \sum_{m=-\infty}^{\infty} U_H \left( \frac{m \lambda}{2},0 \right) \frac{ \exp (i k R_m)}{\sqrt{R_m}} \cos \theta_m
\end{equation}
where $R_m$ is the distance between the points at $(x,z)$ and $(m\lambda/2,0)$, and $\theta_m$ is the angle between the ray connecting these two points and the $z$-axis as illustrated in Fig. 1.
\begin{figure}[h]
\centering
\includegraphics[width=0.3\textwidth]{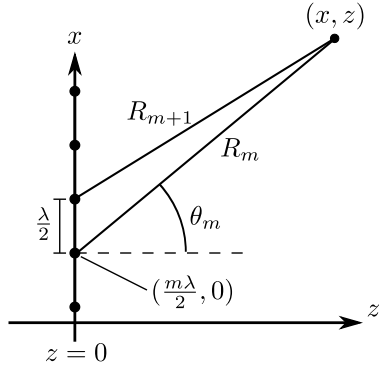}
\caption{A two-dimensional electromagnetic disturbance, composed solely of homogeneous plane waves and satisfying the given boundary conditions, is uniquely determined at any point $(x,z)$ in the half-space $z \geq 0$ by its values on the boundary $z=0$ at points separated by $\lambda$/2. The field is constructed by placing individual oscillating dipoles at each of these boundary points and adding up all contributions at the point of interest.}
\end{figure}

To the knowledge of the author, there is no exact closed-form expression for the kernel function $G(x,z)$ in terms of elementary functions for the two-dimensional case.
\newpage

\section{Diffraction in Three Dimensions}
We now examine the more general case of diffraction in three dimensions, dealing only with the homogeneous component $U_H$ of the field $U$. Starting from Eq. (7), the evaluation of the integral only requires that the angular spectrum function $A(u,v)$ be defined on the unit disc $u^2+v^2 \leq 1$. Let the function $f(u,v)$ be defined on the square domain, in which both $u$ and $v$ are bound to the interval $[-1,1]$, such that
\begin{equation}
f(u,v) = \left\{
\begin{array}{rl}
A(u,v) & u^2 + v^2 \leq 1 \\
0 & u^2 + v^2 > 1
\end{array} \right\}.
\end{equation}
Within this square domain, the function $f(u,v)$ can be expanded as the two-dimensional Fourier series,
\begin{equation}
f(u,v) = \sum_{m,n = -\infty}^{\infty} c_{m,n} \exp (-i \pi [m u + n v ]).
\label{FourierSer2}
\end{equation}
As before, the expansion coefficients are
\begin{equation}
\begin{array}{rl}
c_{m,n} & = \frac{1}{4} \int_{-1}^{1} \int_{-1}^{1} f(u,v) \exp (i \pi [m u + n v ] ) \ du \ dv \\ \\
		& =  \frac{1}{4} \iint_{u^2+v^2 \leq 1} A(u,v) \exp (i \pi [m u + n v ] ) \ du \ dv \\ \\
		& = \frac{1}{4} U_H \left( \frac{m \lambda}{2}, \frac{n \lambda}{2}, 0 \right)
\end{array}
\label{FourierCoeff2}
\end{equation}
where the third equality follows from the second upon comparing with Eq. (\ref{homo}). Because $f(u,v) = A(u,v)$ within the unit disc, the function $A(u,v)$ can be replaced by $f(u,v)$ in Eq. (\ref{homo}). Combining this with Eqs. (\ref{FourierSer2}) and (\ref{FourierCoeff2}) then yields
\begin{equation}
U_H(x,y,z) = \sum_{m,n = -\infty}^{\infty} U_H \left( \frac{m \lambda}{2},\frac{n \lambda}{2},0 \right) G \left( x-\frac{m \lambda}{2},y-\frac{n \lambda}{2}, z \right)
\label{discretehomo2}
\end{equation}
where
\begin{equation}
G(x,y,z) = \frac{1}{4} \iint_{u^2+v^2 \leq 1} \exp (ik [ux + vy + wz]) \ du \ dv.
\label{G}
\end{equation}
is the three dimensional kernel function.

To evaluate the integral in Eq. (\ref{G}), we make use of the coordinate transformations illustrated by Fig. 2 and described as follows. The approach is similar to that in Ref. \cite{Weyl:1919ys}, but here a more careful transformation of the limits of integration is required. We first make the transformation $(x,y,z) \rightarrow (x',y',z')$, consisting of a rotation about the origin, such that the point $P$ with coordinates $(x,y,z)$ lies on the $z'$-axis in the new coordinate system. That is, $x'=y'=0$ for point $P$. The same transformation is also applied to the direction cosines, $(u,v,w) \rightarrow (u',v',w')$. This change of variables preserves the dot product,
\begin{equation}
ux+vy+wz = u'x'+v'y'+w'z'= w' r
\end{equation}
where $r = \sqrt{x^2+y^2+z^2}$. The ratio of the differential areas $du \ dv$ and $du' \ dv'$ (given by the determinant of the Jacobian) is $\cos \theta$, where $\theta = \cos^{-1} (z/r)$ is the angle between point $P$ and the $z$-axis. With this,
\begin{equation}
\frac{1}{4} \iint \exp (ik [ux + vy + wz]) \ du \ dv = \frac{\cos \theta}{4} \iint \exp (ik w' r) \ du' \ dv'.
\end{equation}
This form suggests a switch to spherical coordinates in the space of the direction cosines,
\begin{equation}
\begin{array}{rl}
u' &= \ \sin \eta \ \cos \psi \\
v' &= \ \sin \eta \ \sin \psi  \\
w' &= \ \cos \eta
\end{array}
\end{equation}
where $\eta$ and $\psi$ are the azimuthal and polar angles, respectively, of the point with coordinates $(u',v',w')$. The differential area transforms once more as 
\begin{equation}
du' \ dv' = \sin \eta \ \cos \eta \ d \eta \ d \psi.
\end{equation}
The limits of integration, defined by $u^2+v^2 \leq 1$ in the initial coordinate system, transform accordingly. The polar angle $\psi$ is integrated from 0 to $2\pi$. The azimuthal angle $\eta$ is integrated from 0 to $\pi/2 + \Delta \eta$, where $\Delta \eta$ is the angle between the planes $z=0$ and $z'=0$ as seen from the origin at angle $\psi$. Applying the spherical law of sines \cite{Zwillinger:2003fk} to the geometry shown in the inset of Fig. 2, it can be easily verified that $\Delta \eta =  \sin^{-1} (\cos \psi \sin \theta)$. Therefore, the integral in Eq. (\ref{G}) becomes
\begin{equation}
\begin{array}{rl}
\frac{1}{4} \iint_{u^2+v^2 \leq 1} \exp (ik [ux + vy + wz]) \ du \ dv &= \frac{\cos \theta}{4ik} \frac{d}{dr} \int_{0}^{2\pi} \int_{0}^{\frac{\pi}{2}+ \sin^{-1} (\cos \psi \sin \theta)} \exp (ikr \cos \eta) \sin \eta \ d \eta \ d \psi \\ \\
												&= \frac{1}{4ik} \frac{d}{dz} \int_{0}^{2\pi} \int_{0}^{\frac{\pi}{2}+ \sin^{-1} (\cos \psi \sin \theta)} \exp (ikr \cos \eta) \sin \eta \ d \eta \ d \psi.
\end{array}
\end{equation}
The chain rule is applied, $\cos \theta \ (d/dr) = (dr/dz) (d/dr)  = d/dz$, to obtain the second equality.
\begin{figure}[h]
\centering
\includegraphics[width=0.4\textwidth]{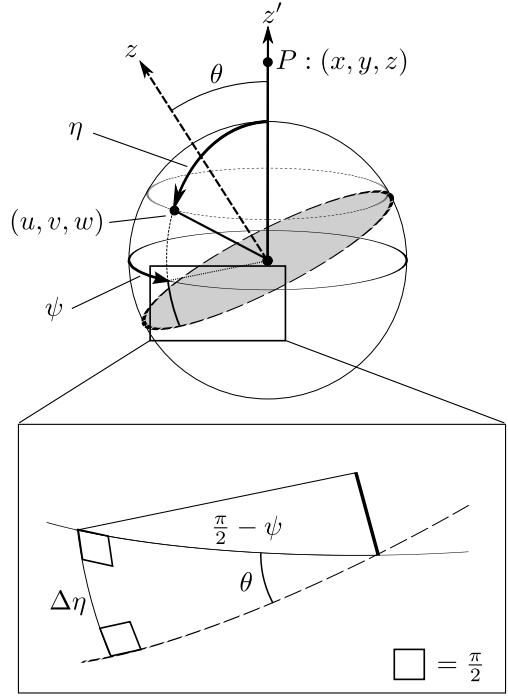}
\caption{Geometry of the coordinate transformation. The point $P$ in the new coordinate system lies on the $z'$-axis. The point given by coordinates $(u,v,w)$ lies on the unit half-sphere $w \geq 0$. The shaded region is the plane $z=0$. In the inset, right angles of the spherical triangle are denoted by small squares.}
\end{figure}
The integral on the right side of Eq. (37) can now be evaluated with the substitution $\tau = \cos \eta$, yielding
\begin{equation}
\begin{array}{rl}
\frac{1}{4} \iint_{u^2+v^2 \leq 1} \exp (ik [ux + vy + wz]) \ du \ dv & = \frac{1}{4ik} \frac{d}{dz} \int_{0}^{2\pi} \int_{-\sin \theta \cos \psi}^{1} \exp (ikr \tau) \ d \tau \ d \psi \\ \\
												& = \frac{1}{4ik} \frac{d}{dz} \int_{0}^{2\pi} \frac{\exp (ikr) - \exp (-ikr \sin \theta \cos \psi )}{ikr} \ d \psi \\ \\
\end{array}
\label{Gint}
\end{equation}
and finally,
\begin{equation}
G(x,y,z) = -\frac{1}{2 \pi} \left( \frac{\lambda}{2} \right)^2 \frac{d}{dz} \frac{\exp (ikr) - J_0 (kr \sin \theta)}{r}.
\label{G2}
\end{equation}
where $J_0$ is the zeroth order Bessel function of the first find, which admits the identity \cite{Abramowitz:1964zr},
\begin{equation}
J_0(s) = \frac{1}{2\pi} \int_{0}^{2\pi} \exp (i s \cos \phi) \ d \phi,
\end{equation}
used to obtain Eq. (\ref{G2}) from Eq. (\ref{Gint}).
Note that, whereas we were only able to obtain an approximate simplification of the kernel function $G$ in the two-dimensional case, Eq. (\ref{G2}) is an exact expression of this function for three dimensions. Applying this to Eq. (\ref{discretehomo2}) gives the discretized diffraction formula,
\begin{equation}
U_H(x,y,z) = -\frac{1}{2 \pi} \left( \frac{\lambda}{2} \right)^2 \sum_{m,n = -\infty}^{\infty} U_H \left( \frac{m \lambda}{2},\frac{n \lambda}{2},0 \right)  \frac{d}{dz} \frac{\exp (ikR_{m,n}) - J_0 (kR_{m,n} \sin \theta_{m,n})}{R_{m,n}}
\label{DiscDiff}
\end{equation}
where $R_{m,n} = \sqrt{\left(x-m \lambda/2 \right)^2+\left(y-n \lambda/2\right)^2+z^2}$ is the distance between the point at $(m \lambda / 2,n \lambda/2,0)$ and point $P$, and $\theta_{m,n} = \cos^{-1} (z/R_{m,n}) $ is the angle between the ray connecting these two points and the $z$-axis, as illustrated in Fig. 3. Equation (\ref{DiscDiff}) bears much resemblance to the Rayleigh-Sommerfeld integral formula of Eq. (\ref{Rayleigh}), with the exception of the Bessel function term, and more importantly the integral is replaced by a sum.
\begin{figure}[ht]
\centering
\includegraphics[width=0.4\textwidth]{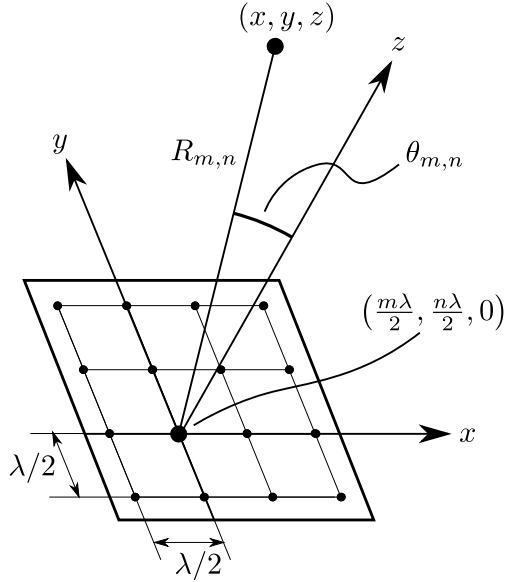}
\caption{The homogeneous component $U_H$ of the diffracted field at the point $(x,y,z)$ in the half-space $z \geq 1$ is completely determined by its values on the rectangular lattice of points at $(m \lambda /2,n \lambda / 2,0)$, where $m$ and $n$ are integers. It is found by placing secondary sources at each lattice point and adding up each contribution at the point of interest.}
\end{figure}

\newpage
\section{Discussion and Conclusions}
We present a solution to the scalar Helmholtz equation, the cornerstone of Scalar Diffraction Theory, which inherently implies a potentially useful sampling theorem regarding the propagation of electromagnetic radiation in free space. Summarily, if the scalar field is composed only of homogeneous plane waves, then it can be perfectly reconstructed from the knowledge of its values on a lattice of points on the boundary $z=0$. The lattice of points has uniform spacing of $\lambda / 2$ between collinear points. If the field is composed of inhomogeneous plane waves, then one can always choose some point $z$ such that this contribution to the total field becomes negligibly small.

Going further, the solution presented in Eq. (\ref{DiscDiff}) can be straightforwardly interpreted as a matrix equation, if the field were to be evaluated on a secondary lattice of points for some $z>0$. Posing the diffraction calculation in this form would be well-suited for numerical algorithms performed by a computer. The operation could possibly be optimized for efficiency, analogous to the Fast Fourier Transform. Such a development would significantly reduce the time needed to calculate the diffracted field in many applications.

\section{Acknowledgements}
I would like to thank Wayne R. McKinney of Lawrence Berkeley National Laboratory for several valuable discussions on this topic.


\begin{thebibliography}{10}
\newcommand{\enquote}[1]{``#1''}

\bibitem{Goodman:1988fk}
J.~W. Goodman, \emph{Introduction to Fourier Optics} (McGraw-Hill, 1988),
  chap.~3, 2nd ed.

\bibitem{Born:1980uq}
M.~Born and E.~Wolf, \emph{Principles of Optics} (Pergamom Press, 1980),
  chap.~8, 6th ed.

\bibitem{Goodman:1988kx}
J.~W. Goodman, \emph{Introduction to Fourier Optics} (McGraw-Hill, 1988),
  chap.~4, 2nd ed.

\bibitem{Hecht:2002fk}
E.~Hecht, \emph{Optics} (Addison Wesley, 2002), chap.~10, 4th ed.

\bibitem{Mandel:1995fk}
L.~Mandel and E.~Wolf, \emph{Optical Coherence and Quantum Optics} (Cambridge
  University Press, 1995), chap.~3.

\bibitem{Boas:2006uq}
M.~L. Boas, \emph{Mathematical Methods in the Physical Sciences} (John Wiley
  and Sons, 2006), chap.~7.

\bibitem{Whittaker:1915kx}
E.~T. Whittaker, \enquote{On the functions which are represented by the
  expansions of the interpolation theory,} Proc. Royal Soc. Edinburgh
  \textbf{35}, 181--194 (1915).

\bibitem{Whittaker:1927vn}
J.~M. Whittaker, \enquote{On the cardinal function of interpolation theory,}
  Proc. Edinburgh Math. Soc. \textbf{1}, 41--46 (1927).

\bibitem{Shannon:1948uq}
C.~E. Shannon, \enquote{A mathematical theory of communication,} The Bell
  System Technical Journal \textbf{27}, 379--423 (1948).

\bibitem{Goodman:1988fj}
J.~W. Goodman, \emph{Introduction to Fourier Optics} (McGraw-Hill, 1988),
  chap.~2, 2nd ed.

\bibitem{Erdelyi:1956zr}
A.~Erd{\'e}lyi, \emph{Asymptotic Expansions} (Dover, reprint, 1956), chap.~2.

\bibitem{Weyl:1919ys}
H.~Weyl, \enquote{Ausbreitung elektromagnetischer wellen uber einem ebenem
  leiter,} Ann. Phys. \textbf{60}, 481--500 (1919).

\bibitem{Zwillinger:2003fk}
D.~Zwillinger, \emph{Standard Mathematical Tables and Formulae} (CRC Press,
  2003), chap.~6.

\bibitem{Abramowitz:1964zr}
M.~Abramowitz and I.~A. Stegun, \emph{Handbook of Mathematical Functions}
  (National Bureau of Standards, 1964), chap.~9, 10th ed.

\end{thebibliography}

\bibliographystyle{osajnl}

\end{document}